\documentclass[%
reprint,
amsmath,amssymb,
aps,
superscriptaddress,
preprintnumbers
]{revtex4-2}

\usepackage{graphicx}
\usepackage{dcolumn}
\usepackage{bm}
\usepackage{hyperref}

\usepackage{color}
\usepackage{cancel}
\usepackage{AAS_macro}

\newcommand*\df{\mathop{}\!\mathrm{d}}
\newcommand{\n}{\nonumber}

\DeclareMathOperator{\sgn}{sgn}

\makeatletter
\newcommand*{\rom}[1]{\expandafter\@slowromancap\romannumeral #1@}
\makeatother

\usepackage{multirow}

\begin{document}

\title{Stochastic Fluctuations of Low-Energy Cosmic Rays\\ and the Interpretation of Voyager Data}%

\author{Vo Hong Minh Phan}
\email{vhmphan@physik.rwth-aachen.de}
\affiliation{Institute for Theoretical Particle Physics and Cosmology (TTK), RWTH Aachen University, 52056 Aachen, Germany}
\affiliation{Universit\'e de Paris, CNRS, Astroparticule et Cosmologie, F-75006 Paris, France}

\author{Florian Schulze}
\email{florian.tobias.schulze@rwth-aachen.de}
\affiliation{Institute for Theoretical Particle Physics and Cosmology (TTK), RWTH Aachen University, 52056 Aachen, Germany}

\author{Philipp Mertsch}
\email{pmertsch@physik.rwth-aachen.de}
\affiliation{Institute for Theoretical Particle Physics and Cosmology (TTK), RWTH Aachen University, 52056 Aachen, Germany}

\author{Sarah Recchia}
\email{sarah.recchia@unito.it}
\affiliation{Dipartimento di Fisica, Universit\'a di Torino, via P. Giuria 1, 10125 Torino, Italy}

\author{Stefano Gabici}
\email{gabici@apc.in2p3.fr}
\affiliation{Universit\'e de Paris, CNRS, Astroparticule et Cosmologie, F-75006 Paris, France}

\date{\today}

\preprint{TTK-21-14}

\begin{abstract}
Data from the Voyager probes have provided us with the first measurement of cosmic ray intensities at MeV energies, an energy range which had previously not been explored. 
Simple extrapolations of models that fit data at GeV energies, e.g. from AMS-02, however, fail to reproduce the Voyager data in that the predicted intensities are too high. 
Oftentimes, this discrepancy is addressed by adding a break to the source spectrum or the diffusion coefficient in an \textit{ad hoc} fashion, with a convincing physical explanation yet to be provided. 
Here, we argue that the discrete nature of cosmic ray sources, which is usually ignored, is instead a more likely explanation. 
We model the distribution of intensities expected from a statistical model of discrete sources and show that its expectation value is not representative, but has a spectral shape different from that for a typical configuration of sources.
The Voyager proton and electron data are however compatible with the median of the intensity distribution.
\end{abstract}

\maketitle

\section{Introduction}
Despite over 100 years of intense experimental and theoretical efforts, the origin of Galactic cosmic rays (GCRs) has still not been unambiguously identified. At energies above a few tens of GeV, much progress has been made in the last couple of years, thanks to direct observations by high-precision, high-statistics experiments like AMS-02 or PAMELA and the study of gamma-rays by \textit{Fermi}-LAT and Cherenkov telescopes~\cite{gabici2019}. At lower energies, however, the situation is still very much unclear. Until recently, solar modulation, that is the suppression of intensities due to interactions with the magnetised solar wind, hampered the study of GCRs at energies around a GeV and below~\cite{potgieter2013}. Modelling of the transport of these particles therefore essentially relied on extrapolations from higher energies.

In 2013, however, the first direct observations of interstellar spectra by Voyager~1 were published and it became clear that simple extrapolations from higher energies fail~\cite{stone2013}. Specifically, in order to fit both Voyager~1 and \mbox{AMS-02} data, simple diffusive transport models overpredict the intensities at Voyager energies (e.g.~\cite{Vittino:2019yme}). While phenomenological models can add a break in the source spectra around a GeV in an \emph{ad hoc} fashion, the physical interpretation of such a break is rather questionable~\cite{cummings2016,orlando2018,boschini2018a,boschini2018b,johannesson2018,bisschoff2019}. In fact, we would maintain that no convincing explanation of such a break has been put forward to date.

This issue is far from academic since the energy range affected is important for a number of issues. 
In fact, most of the energy density of GCRs is contributed in the energy range around a GeV and, depending on the spectrum, possibly below.
Correspondingly, different spectra imply different power requirements for the sources, which provide helpful clues on the nature of GCR acceleration \cite{ginzburg1964,recchia2019}. Moreover, GCRs are the prime agent of ionisation in dense molecular clouds (MCs) and recently, the ionisation rates inferred from nearby MCs have been shown to be in strong tensions with the local interstellar spectra as measured by Voyager~1 \cite{phan2018,silsbee2019,padovani2020}. Furthermore, diffuse emission in radio waves and MeV gamma-rays is sensitive to this energy range (e.g.~\cite{orlando2018}). The diffuse radio background constitutes the dominant foreground for upcoming cosmological studies of the epoch of reionisation (e.g. \cite{Rao:2016xre}) and diffuse gamma-rays for proposed MeV missions (eAstrogam~\cite{DeAngelis:2016slk}, AMEGO~\cite{2019BAAS...51g.245M}). Lastly, the current picture of GCRs is simply incomplete if one cannot explain cosmic rays at MeV energies.

An important effect for MeV GCRs that has been ignored in the literature is due to the discrete nature of sources. Instead, the distribution of sources in position and time is oftentimes modelled as smooth. That is, the predicted cosmic ray density $\psi$ is the solution of the transport equation with a source term $q$ that is a smooth function of position ($r$ and $z$), energy $E$ and time $t$,
\begin{equation}
\frac{\partial \psi}{\partial t}+\frac{\partial}{\partial z}\left(u \psi \right) -D\nabla^2 \psi + \frac{\partial }{\partial E}\left(\dot{E}\psi\right)=q(r, z, E, t) \, . \label{eq:transport}
\end{equation}
Here, $u=u(z)$ is the advection velocity profile with only the component perpendicular to the Galactic disk, $D=D(E)$ is the isotropic and homogeneous diffusion coefficient, $\dot{E}$ describes the energy loss rate for GCRs both inside the Galactic disk and in the magnetized halo. Note that it might be more customary to formulate Eq. 1 in terms of momentum (see \footnote{See Supplemental Material at \url{http://link.aps.org/supplemental/} for some discussions with supporting figures and tables, which includes Refs. \cite{strong1998,schlickeiser1999,mertsch2020}.} for the transformation to kinetic energy).

Even though the sources are likely separate, discrete objects like supernova remnants (SNRs), the approximation of a smooth source density is admissible at GeV energies, since the transport distances and times exceed the typical source separations and ages. However, if energy losses reduce the propagation times and distances, this approximation breaks down and instead the discrete nature of the sources needs to be taken into account. This can be done by replacing the smooth source density from before by a sum of individual delta-functions in distance and age,
\begin{equation}
q(r, z, E, t) = \sum_{i=1}^{N_\text{s}} Q(E)\frac{\delta(r - r_i)}{2\pi r_i}\delta(z-z_i)\delta(t - t_i) \, .
\end{equation}
$Q(E)$ denotes the spectrum that an individual source injects into the ISM. The total intensity from $N_{\text{s}}$ sources is then just the sum over the Green's function $\mathcal{G}(r, z, E; r_i, z_i, t-t_i)$ of Eq.~\eqref{eq:transport} at the position of the solar system,
\begin{equation}
\psi = \sum_i \mathcal{G}(r=0, z=z_\odot, E; r_i, z_i, t-t_i) \, . \label{eq:stochasticity}
\end{equation}
where $z=z_{\odot}\simeq 14$ pc is the vertical offset of the solar system from the Galactic mid-plane \cite{skowron2019}. An example where this approach has been followed are high-energy electrons and positrons at hundreds of GeV and above, which lose energy due to the synchrotron and inverse Compton processes~\cite{mertsch2011}, but ionisation losses also severely limit the propagation of MeV GCRs. Predicting their local intensities therefore requires rather precise knowledge of the ages and distances of the sources. While some young and nearby sources might be known, catalogues of such sources remain necessarily incomplete, in particular with respect to far away and old sources.

Instead, the distribution of sources can be considered a statistical ensemble, thus opening the path towards a statistical modelling of GCR intensities. Operationally, one draws a set of source distances and ages from the statistical probability density function (PDF). Adding up their intensities results in a prediction for this given realisation of the sources. Repeating this procedure for a large number of realisations, one can estimate the distribution of intensities. The first moment and second central moment of this distribution are the expectation value and the variance. Since the expectation value $\langle \psi \rangle$ could be obtained by averaging over many realizations, it approaches the solution of the GCR transport equation~\eqref{eq:transport} when the smooth source PDF, from which individual source distance and ages are drawn, is used as the source term $q$. However, as it turns out the statistics of the intensities is markedly non-Gaussian, with the second moment divergent. This is due to the long power-law tails of the intensity PDF. Its asymmetric shape renders the expectation value different from the median and from the maximum of the distribution \cite{nolan2020}.

In this \emph{letter}, we model the intensities of GCR protons and electrons between $1 \, \text{MeV}$ and $10 \, \text{GeV}$ taking into account the stochasticity induced by the discreteness of sources. Consequently, our predictions will be probabilistic. We will illustrate that the expectation value is a bad estimator for the intensities in individual realisations. For instance, for low enough energies the expectation value is outside the $68\%$ uncertainty band. Furthermore, its spectral shape is markedly different than the intensity in any individual realisation. Finally, we stress that the expectation value does not reproduce the data either unless an artificial break is added to the source spectrum. Instead, we suggest considering the median of the intensity PDF as a better measure of what a ``typical'' intensity will look like, and the reference intensity around which the intensities from all realisations are distributed. Interestingly, the data for protons and electrons fall squarely within the uncertainty bands. We thus conclude that a model without artificial breaks is to be preferred in explaining the Voyager~1 and AMS-02 data as long as the stochasticity effect is taken into account.

\section{Modelling}
\label{sec:stochastic}

Equation \ref{eq:transport} is solved numerically assuming GCRs propagate within a finite cylindrical region with height $2L\simeq 8$ kpc and radius $r_{max}\simeq$ 10 kpc centering around the source. The other parameters of our model are chosen such that the most probable values of the intensity is compatible with the observational data. Specifically, the advection velocity is assumed to have the following profile $u(z)=u_0\sgn(z)$ with $u_0=16$ km/s, where $\sgn(z)$ is the sign function. We assume also the diffusion coefficient of the form $D(E)\sim \beta\gamma^{\delta}$ as suggested in \citep{schlickeiser2010} where $\beta=v/c$ is the ratio between the particle's speed and the speed of light and $\gamma$ is the particle's Lorentz factor (note that assuming the diffusion coefficient to scale with rigidity might not qualitatively alter the results \cite{Note1}). 
Recent analyses of GCRs seem to suggest slightly different values for $\delta$ depending on whether or not the unstable isotope $^{10}$Be is taken into account \cite{evoli2019,evoli2020,weinrich2020}. However, the overall results of the local spectra would remain qualitatively unchanged for different values of $\delta$ if we slightly modified the injection spectra. In the following, we shall adopt $\delta=0.63$ and normalize the diffusion coefficient such that $D(E=10\textrm{ GeV})\simeq 5\times 10^{28}$ cm$^2$/s for both species \citep{evoli2019}. We caution that the diffusion coefficient in the disk and in the halo could in principle be different and so our parametrisation is to be regarded as a suitably defined average. 

Low-energy GCRs lose energy mostly due to ionisation interactions with the neutral gas in the disk as discussed above. There are also proton-proton interactions and radiative energy loss at high energies. All the energy loss mechanisms are effective only within the disk of size $2h\simeq 300$ pc apart from synchrotron and inverse Compton processes. More importantly, the rate of energy loss depends also on the average number density of the hydrogen atoms in the disk. We adopt $n_\text{H}=0.9$ cm$^{-3}$ corresponding to the surface density of 2 mg/cm$^{2}$, which is roughly the observed value \citep{ferriere2001}. The specific form of the energy loss rate are collected from \citep{schlickeiser2002,mertsch2011,krakau2015,evoli2017} (see also \cite{Note1}). 

We take into account also the adiabatic energy loss due to advection with the approximation $|\dot{E}_{ad}|=2pv u_0\delta(z) \simeq pv u_0/(3h)$ \cite{jaupart2018}. As for the injection spectrum, we shall adopt the following power-law form in momentum down to the kinetic energy of 1 MeV:
\begin{eqnarray}
Q(E)=\frac{\xi_{CR}E_{SNR}}{(mc^2)^2\Lambda\beta}\left(\frac{p}{mc}\right)^{2-\alpha},\label{eq:source_function}
\end{eqnarray} 
where $\xi_{CR}=8.7\%$ and $\xi_{CR}=0.55\%$ are the acceleration efficiencies of the source for GCR protons and electrons respectively, $E_{SNR}\simeq 10^{51}$ erg is the total kinetic energy of the supernova explosion, $m$ is the mass of the GCR species of interest, and
\begin{eqnarray}
\Lambda=\int^{p_{max}}_{p_{min}}\left(\frac{p}{mc}\right)^{2-\alpha}\left[\sqrt{\left(\frac{p}{mc}\right)^2+1}-1\right]\frac{\df p}{mc}.\label{eq:Lambda_Q}
\end{eqnarray}
We shall take $\alpha=4.23$ as suggested for the fit at high energies \cite{evoli2019}. Such a power-law in momentum seems to be preferred from the commonly accepted theory of diffusive acceleration on SNR shocks \citep{malkov2001,blasi2013}. Even though the extension of the spectrum down to 1 MeV seems questionable, there exist observational evidences of enhanced ionisation rates in the vicinity of SNRs indicating the presence of low-energy GCRs accelerated from these objects \citep{vaupre2014,gabici2015,phan2020}. Note that we neglect stochastic re-acceleration for simplicity and this process might be examined in future works. 

\begin{figure*}[htpb]
\centerline{
\includegraphics[width=3.5in, height=2.8in]{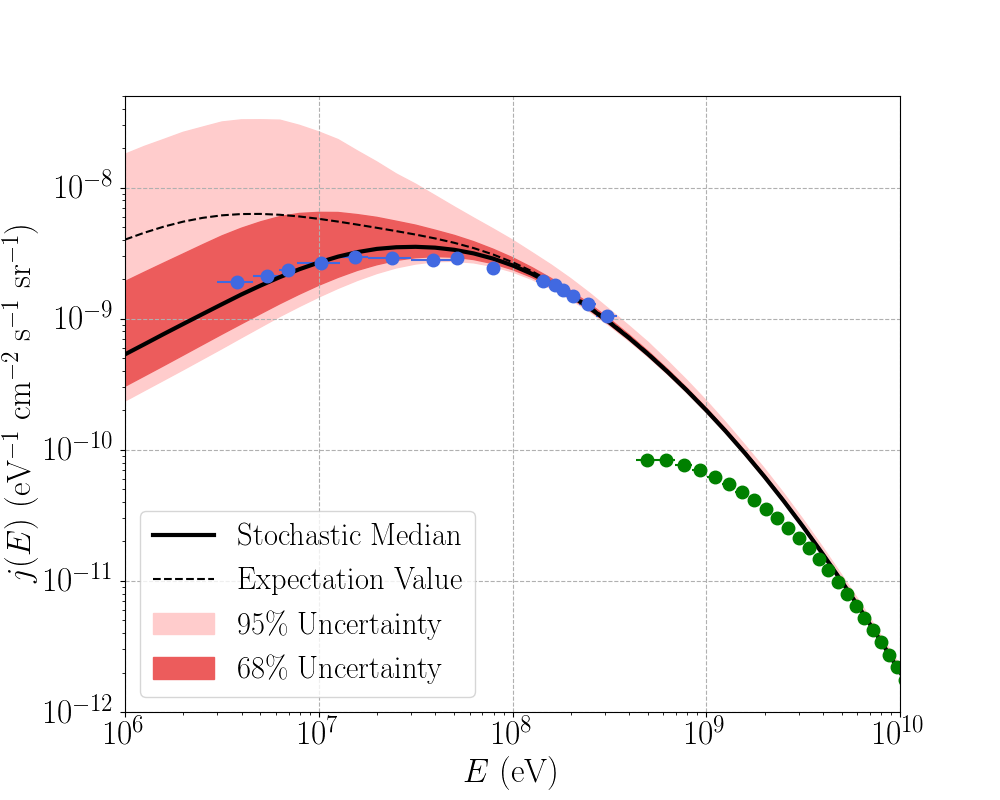}
\includegraphics[width=3.5in, height=2.8in]{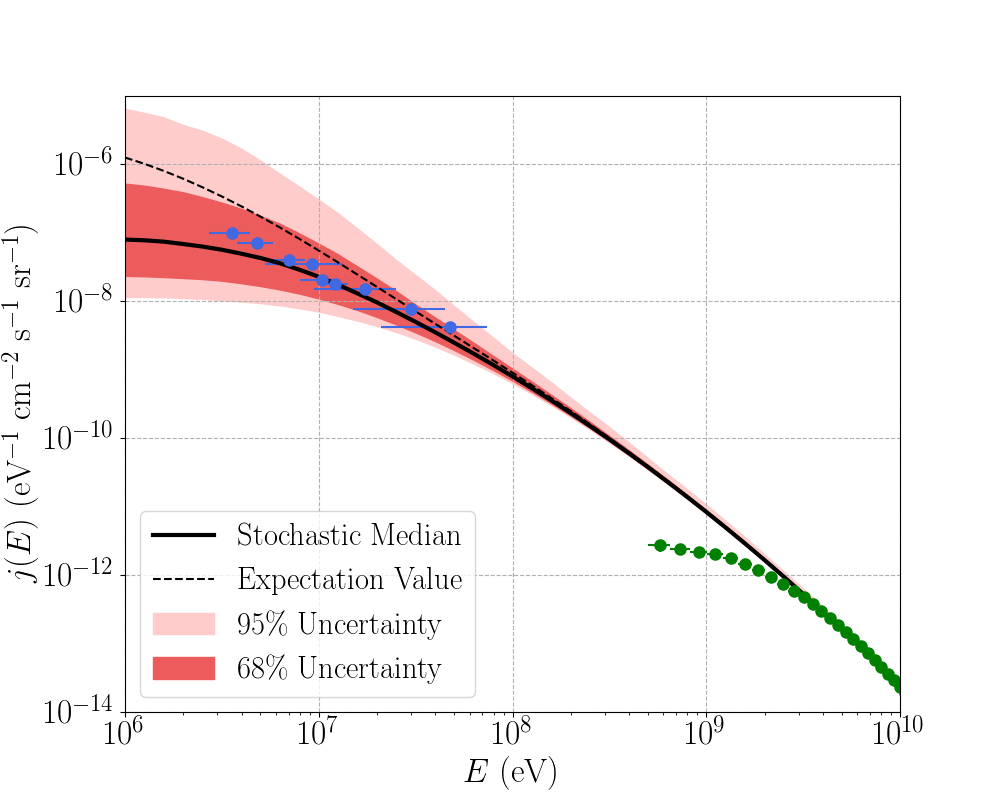}
}
\caption{Stochastic fluctuations of GCR protons (left panel) and electrons (right panel) in comparison with data from Voyager~1~\cite{cummings2016} (blue) and AMS-02~\cite{AMS2014,AMS2015} (green). The dotted and solid black curves are respectively the expectation values and the median of the intensities. The shaded regions are the 95\% and 68\% uncertainty ranges.}
\label{fg:stochastic}
\end{figure*}

We have built up a statistical ensemble by generating a large number $N_\text{r}=2000$ of realisations, in each drawing a large number of sources $N_\text{s}$ from the spatial distribution following a spiral pattern \citep{vallee2005} with a radial modulation \citep{case1998}, as employed in~\cite{mertsch2011}, and with a homogeneous distributions for the time since injection and for the vertical position of sources. We limit ourselves to $r_i^{(n)}< r_{max}=10$ kpc and the time since injection $\tau_{i}^{(n)}<\tau_{max}=10^8$ yr since older and further sources would not contribute significantly. The total number of discrete sources in each realisation could be estimated roughly as $N_{s}=\mathcal{R}_{s}\tau_{max}r_{max}^2/R_d^2\simeq 1.33\times 10^6$, where $\mathcal{R}_{s}\simeq 0.03$ yr$^{-1}$ is the source rate and $R_d\simeq 15$ kpc is the radius of the Galactic disk. We adopt $2h_s\simeq 80 \, \text{pc}$ for the vertical extension of sources expected for CCSN \citep{prantzos2011}.

We thus obtain an ensemble of intensities \mbox{$j^{(n)} = v/(4 \pi) \psi^{(n)}$} for the individual source realisations $n$ that we can characterise statistically. For instance, a histogram of these intensities at a specific energy could serve as an estimate of the intensity PDF $p(j)$. Note that the expectation value of the intensity $\langle j \rangle = \int \mathrm{d} j \, p(j)$ is equal to the intensity predicted for the smooth source density of Ref.~\cite{mertsch2011}. We have found $p(j)$ to be extremely non-Gaussian with power-law tails, e.g. $p(j) \propto j^{-2}$ for $j \gg \langle j \rangle$ at $E=1$ MeV. In fact, these distribution functions are not only asymmetric but they also do not have a well-defined second moment as shown for similar analyses at high energies \citep{mertsch2011,blasi2012,bernard2012,genolini2017}. We shall, therefore, specify the uncertainty intervals of the intensity using the percentiles as in \citep{mertsch2011}, e.g. $j_{a\%}$ is defined via $a\%=\int_0^{j_{a\%}} \df j \, p(j)$. The $68$\% and $95\%$ uncertainty range of the intensity $j(E)$ are then $\mathcal{I}_{68\%}=\left[j_{16\%},j_{84\%}\right]$ and $\mathcal{I}_{95\%}=\left[j_{2.5\%},j_{97.5\%}\right]$.

\section{Results and Discussion}
\label{sec:results}

We present in Fig. \ref{fg:stochastic} the $95\%$ and $68\%$ uncertainty bands of the intensities for both GCR protons (left panel) and electrons (right panel) in the energy range from 1 MeV to about 10 GeV together with the expectation values of the intensities and data from Voyager~1~\citep{cummings2016} and AMS-02~\citep{AMS2014,AMS2015}. The uncertainty ranges above 100 MeV are quite narrow since the energy loss time and the diffusive escape time are sufficiently large such that the distribution of GCRs inside the Galactic disk become more or less uniform. We note that this will not remain true for GCR electrons of energy above 10 GeV since the energy loss rate for these particles become increasingly larger in this energy range which will result in significant stochastic fluctuations \citep{atoyan1995,ptuskin2006,mertsch2011,mertsch2018,recchia2019b,manconi2020,evoli2021}.

\begin{figure}[ht]
\includegraphics[width=3.5in, height=2.8in]{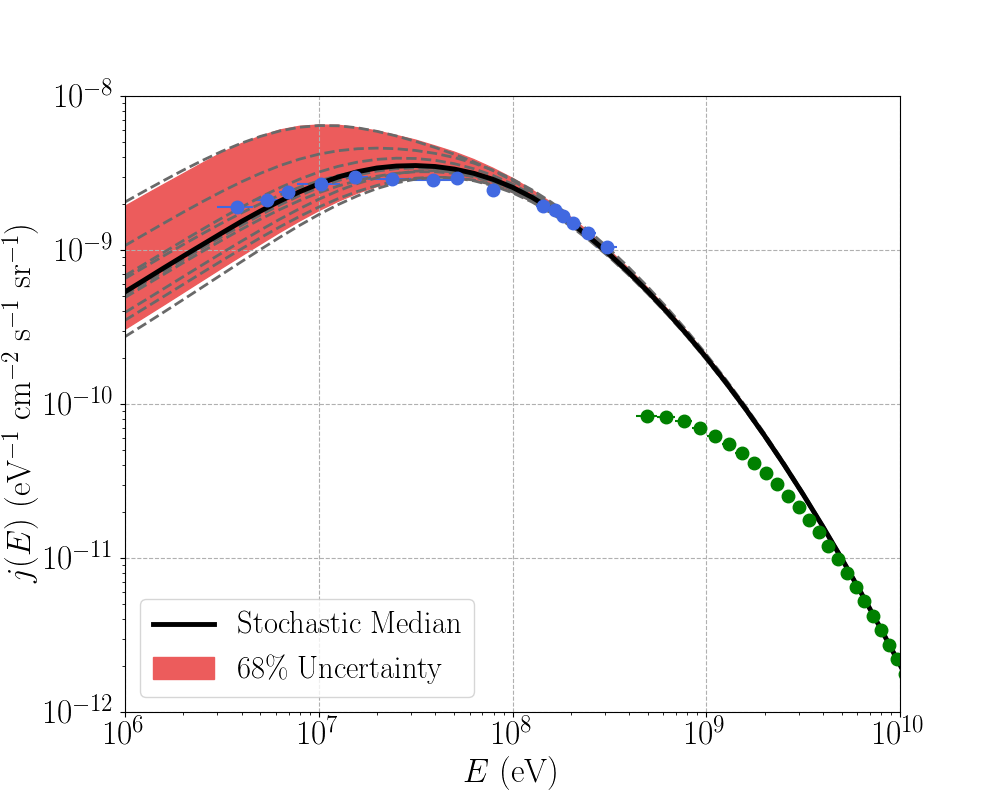}
\caption{Intensities of GCR protons for several realizations (dashed grey curves) around the 68\% uncertainty range (shaded region). Data points are as in Fig. \ref{fg:stochastic} and the solid black curve is the median of the intensities.}
\label{fg:sample}
\end{figure}

The uncertainty ranges broaden for $E\lesssim100$ MeV until a characteristic energy $E^*$ below which the ratio between the upper and lower limit of the intensities becomes constant. 
Such a feature emerges from the fact that the Green's function behaves as \mbox{$\mathcal{G}(r=0,z=z_\odot,E,r_i,z,z_i,\tau_i)\sim 1/|\dot{E}|$} if the propagation time $\tau_i$ is much larger than the energy loss time ($\tau_i\gg \tau_l(E)= E/|\dot{E}|$) which is easily fulfilled for particles of energy below a few tens of MeV. Since $\tau^{(n)}_i\gtrsim \tau_l(E\lesssim 10 \textrm{ MeV})$ for $i=\overline{1,N_{s}}$ in each of the $n$th realization, we expect from Eq. \ref{eq:stochasticity} that $j^{(n)}(E)\sim v/|\dot{E}|$ for all realizations at sufficiently low energies and, thus, the limits of the uncertainty ranges should become parallel below a characteristic energy. The intensities of GCR protons for several realizations which are within the 68\% uncertainty range are depicted in Fig.~\ref{fg:sample} to better illustrate the spectral behaviour at low energies. 

Note that a uniform distribution of GCRs will be attained if the number of sources within the diffusion loss length $l_d(E)=\sqrt{4D(E)\tau_l(E)}$ in the disk is much larger than one,
\begin{eqnarray}
\mathcal{R}_s\tau_l(E) \frac{2 l_d^3(E)}{3 R_d^2 h_s}\gg 1 \, .
\end{eqnarray}
The characteristic energy $E^*$ could be estimated by setting the LHS of the above inequality to one, which gives $E^*\simeq 10$ MeV for both species. 

Interestingly, apart from the deviation in the energy range below a few GeVs due to solar modulation, the median corresponding to $j_{50\%}$, the 50\% percentile of the PDF of the intensities, seems to provide a good fit to the data of Voyager~1 and AMS-02 for both GCR protons and electrons (see Fig.~\ref{fg:stochastic}). We note that both the expectation values and the median do not strictly correspond the intensities of any particular realizations of sources. At low energies, however, the expectation value is dominated by a few, but rather unlikely realisations with extreme intensities such that $j^{(n)}(E)> j_{84\%}(E)$ which are outside of the 68\% uncertainty range. Furthermore, the resulting $\langle j(E)\rangle$, which is also the intensities predicted for the smooth source density as stressed above, has a different energy dependence than the \textit{universal} scaling $j^{(n)}(E)\sim v/|\dot{E}|$ expected at low energies. The median, on the other hand, behaves as $j_{50\%}(E) \sim v/|\dot{E}|$ and, in fact, the intensities in many realizations seems to closely resemble the spectral behaviour of the median both at low and high energies (see Fig. \ref{fg:sample}). It is for this reason that the median is to be preferred over the expectation value for the comparison with observational data. 

We note also that the observed proton spectrum seems to have a broader peak than the median of the stochastic model and the observed electron spectrum seems to exceed the median. It is clear, however, that the local ISM should be quite inhomogeneous, and that the observed spectra in an inhomogeneous ISM could be modelled as the weighted average of spectra for different gas densities to provide better agreement with data. We relegate the details of this to future work.

It is worth mentioning also that the model with the smooth source density could fit data from both Voyager~1 and AMS-02 data under the assumption that the vertical extension of sources is $2h_{s}\simeq 600$ pc \citep{schlickeiser2014} expected for type \rom{1}a SN but these events have a relatively low rate \citep{prantzos2011}. The stochastic model, however, predicts the observational data to be within the most probable range of the intensities for both GCR protons and electrons with $2h_s\simeq 80$ pc comparable to the vertical extension of CCSN with a higher rate \cite{prantzos2011}. More importantly, there is no need to introduce ad hoc breaks both in the injection spectra and the diffusion coefficients. The stochastic model, therefore, seems to be a more appropriate framework for low-energy GCRs.

\section{Summary and outlook}

In this \textit{letter} we have presented results of a modelling of proton and electron spectra between 1 MeV and 10 GeV. Before the advent of the Voyager~1 measurements outside the heliopause, this energy range had received relatively little attention previously due to the fact that solar modulation makes the inference of interstellar spectra difficult. All the models to date assume a smooth source distribution, however, these models do not reproduce the Voyager~1 data unless a spectral break is introduced in the source spectrum. From a microphysical point of view, such a break seems rather unmotivated. 

The smooth approximation is, in fact, not justified since at low energies the energy loss distance becomes shorter than the average source separation. Unlike previous models we therefore considered the discrete nature of sources, modelling the distribution of intensities in a statistical ensemble. We note that the intensity prediction from a smooth density is the ensemble average of this distribution. However, we showed that the ensemble average is not representative of the distribution due to its long power-law tails. For instance, the spectral shapes of the predicted intensities in different realisations are the same below a critical energy. While the expectation value has a very different spectrum at the lowest energies, the median of the distribution does exhibit the same spectral shape. Furthermore, the expectation value is outside the $68\%$ uncertainty range of the distribution at the lowest energies while the median is by definition always inside. We have shown that the Voyager~1 data fall squarely around the median of the distribution without the need for any unphysical breaks in the source spectra (see \cite{Note1} for all model parameters). 

The statistical model we have presented here might have interesting implications for other anomalies observed in low-energy GCRs. For instance, it has been shown recently~\cite{phan2018} that the ionisation rate implied by the Voyager~1 data is much smaller than the ionisation rate directly inferred for a large number of molecular clouds. It would be interesting to see whether the inhomogeneities implied by our statistical model of discrete sources can alleviate this tension. In such a scenario, the Voyager~1 data would need to lie towards the lower edge of the uncertainty band while the molecular cloud measurements would be in regions of systematically higher GCR densities, possibly due to their spatial correlation with source regions. Thanks to our careful statistical model, we will be able to statistically quantify such a model in the future.\\

This project has received funding from the European Union’s Horizon 2020 research and innovation programme under the Marie Skłodowska-Curie grant agreement No 665850. VHMP is grateful to Marco Kuhlen, Nhan Chau, Ngoc Khanh Vu, and Quang Nam Dam for fruitful discussions and technical support.   

\bibliographystyle{apsrev4-2}
\bibliography{mybib}
\newpage

\clearpage
\appendix
\setcounter{equation}{0}
\setcounter{figure}{0}
\widetext
\begin{center}
{\bf \large \large Supplemental Material: Stochastic Fluctuations of Low-Energy Cosmic Rays\\ and the Interpretation of Voyager Data}
\end{center}

%
%
%
%
%
%
%
%

\section*{The cosmic-ray transport equation}
We have adopted the cosmic-ray (CR) transport equation in terms of kinetic energy $E$ for the study of stochasticity. However, it might be more customary to formulate the CR transport equation in terms of momentum $p$. For definiteness, we now lay out the procedure for the transformation. The equation in terms of momentum is \cite{schlickeiser2002}:  
\begin{equation}
\frac{\partial f}{\partial t}+\frac{\partial}{\partial z}\left(u f\right) -D\nabla^2 f + \frac{1}{p^2} \frac{\partial }{\partial p}\left(\dot{p}p^2f\right)=\Tilde{q}(r, z, p, t) \, . \label{eq:transport-p}
\end{equation}
where $f(r,z,p,t)$ is the phase space density of CRs, that is the number of particles per unit volume in configuration and momentum space, $u=u(z)$ is the advection velocity with only component perpendicular to the Galactic disk, $D=D(p)$ is the isotropic and homogeneous diffusion coefficient, and $\dot{p}$ is the momentum loss rate. The phase space density $f(r,z,p,t)$ is related to the cosmic-ray density $\psi(r,z,E,t)$, the number of particles per unit volume and energy, as $\psi(r,z,E,t)=4\pi p^2 f(r,z,p,t)/v$ and, similarly, we have $q(r,z,E,t)=4\pi p^2 \tilde{q}(r,z,p,t)/v$ where $v$ is the particle's speed. It is then clear that we could now transform Eq. \ref{eq:transport-p} into an equation for $\psi(r,z,E,t)$ by performing the change of variable from $p$ to $E$. We note also that $\dot{E}=\dot{p}v$ and, in fact, the standard literature mostly quote the formulae for the energy loss rate (even when Eq. \ref{eq:transport-p} is adopted for the study of CRs \cite{strong1998,schlickeiser2002}).

\section*{Energy loss rate}

Cosmic-ray protons lose energy mostly due to ionization and proton-proton interaction with the gas in the Galactic disk. The combined energy loss rate for these two processes could be written as \cite{schlickeiser2002,krakau2015}: 
\begin{eqnarray}
&&\dot{E}\simeq \textrm{H}(|z|-h) 1.82\times 10^{-7}\left(\frac{n_{\textrm{H}}}{1\textrm{ cm}^{-3}}\right)\n\\
&&\qquad\times\left[(1+0.185\ln\beta)\frac{2\beta^2}{10^{-6}+2\beta^3}+2.115\left(\frac{E}{1\textrm{ GeV}}\right)^{1.28}\left(\frac{E}{1\textrm{ GeV}}+200\right)^{-0.2}\right]\textrm{ eV/s},
\end{eqnarray}
where $H(|z|-h)$ is the Heaviside function which indicates that these energy loss mechanisms are only effective within the height $h$ of the disk, $n_{\mathrm{H}}$ is the density of hydrogen atoms in the disk, $\beta$ is the ratio between the particle's speed and the speed of light, and $E$ is the kinetic energy of the particle.  

Below a few GeV, the main mechanisms for energy loss of CR electrons are ionization interaction and bremsstrahlung radiation in the Galactic disk. At higher energy, these particles lose energy more effectively not only in the disk but also in the CR halo due to synchrotron radiation and inverse Compton scattering. The energy loss rate could then be parametrized as \cite{schlickeiser2002,mertsch2011,evoli2017}:  
\begin{eqnarray}
&&\dot{E}\simeq 10^{-7}\left(\frac{E}{1\textrm{ GeV}}\right)^2+\textrm{H}(|z|-h) 1.02\times 10^{-8}\left(\frac{n_\mathrm{H}}{1\textrm{ cm}^{-3}}\right)\n\\
&&\qquad\qquad\qquad\qquad\qquad\qquad\times\left\{18.495+2.7\ln\gamma+0.051\gamma\left[1+0.137\left(\ln\gamma+0.36\right)\vphantom{^{\frac{^\frac{}{}}{}}}\right]\vphantom{^{\dfrac{}{}}}\right\}\textrm{ eV/s},
\end{eqnarray}
where $\gamma$ is the Lorentz factor.

\section*{Parameters for the stochastic model}
In Tab. \ref{tab:parameters}, we briefly summarise all the parameters adopted in order for the stochastic uncertainty bands to encompass the data from both Voyager 1 and AMS-02. Most of the parameters including the diffusion coefficient and the injection spectra are constrained from the fits at high energies \cite{evoli2019}. 

In fact, the two parameters that only the low-energy spectra are sensitive to are the number density of hydrogen atoms and the advection speed perpendicular to the disk. In fact, the value of the advection speed has also been given in the fits at high energy but it might vary slightly around 10 km/s depending on the model and the species of CRs considered \cite{evoli2019,mertsch2020}. We note also that $n_{\mathrm{H}}$ is not completely free as the surface density of the disk for our Galactic neighborhood is externally constrained to be around 2 g/cm$^{2}$ which is quite consistent with the value adopted for our fits \cite{ferriere2001}. 

\begin{table}[h!]
\centering
\caption{Externally constrained and fitted parameters for the stochastic model for both CR protons and electrons in the case for the diffusion coefficient scaling with Lorentz factor as presented in the main text.}

	\label{tab:parameters}
	\begin{tabular}{|c|c|c|r|} 
		\hline
		\hline
		\multirow{2}{*}{\shortstack{Fitted parameters\\ for low-energy CRs}} & $n_{\mathrm{H}}$ & Gas density in the disk & 0.9 cm$^{-3}$\\
		\cline{2-4}
		& $u_0$ & Advection speed & 16 km/s\\
		\hline
		\multirow{10}{*}{\shortstack{Constrained parameters\\ from high-energy CRs}} & $R_d$ & $\qquad$ Radius of the Galactic disk $\qquad$ & 15 kpc \\
		\cline{2-4}
		& $H$ & Height of the CR halo & 4 kpc\\
		\cline{2-4}
		& $2h$ & Height of the gas disk for energy loss & 300 pc\\
		\cline{2-4}
		& $2h_s$ & Height of the disk of sources & 80 pc\\
		\cline{2-4}
		& $D(E=10\textrm{ GeV})$ & Diffusion coefficient at 10 GeV & $5\times 10^{28}$ cm$^2$/s\\
		\cline{2-4}
		& $\delta$ & Index of the diffusion coefficient & 0.63\\
		\cline{2-4}
		& $\mathcal{R}_s$ & Source rate & 0.03 yr$^{-1}$\\
		\cline{2-4}
		& $\xi_{CR}^{(p)}$ & Proton acceleration efficiency & 8.7\%\\
		\cline{2-4}
		& $\xi_{CR}^{(e)}$ & Electron acceleration efficiency & 0.55\%\\
		\cline{2-4}
		& $\alpha$ & Index of the injection spectra & 4.23\\   
		\hline
		\hline
	\end{tabular}
\end{table}

\begin{figure}[h]
\centerline{
\includegraphics[width=3.2in, height=2.7in]{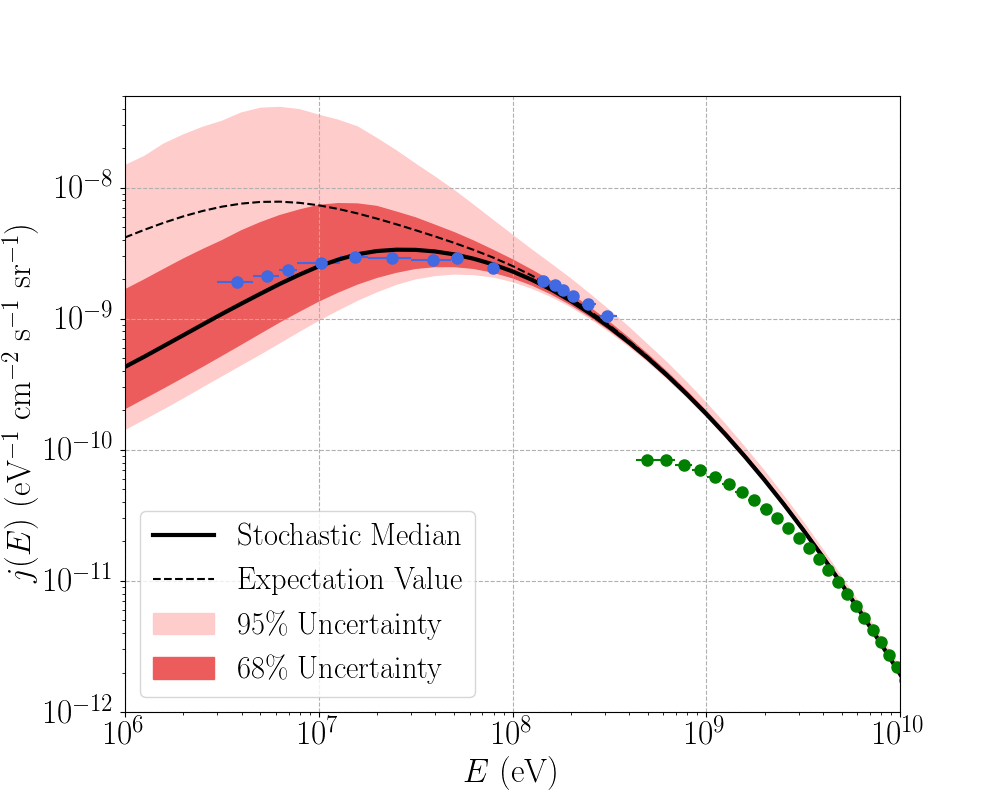}
\includegraphics[width=3.2in, height=2.7in]{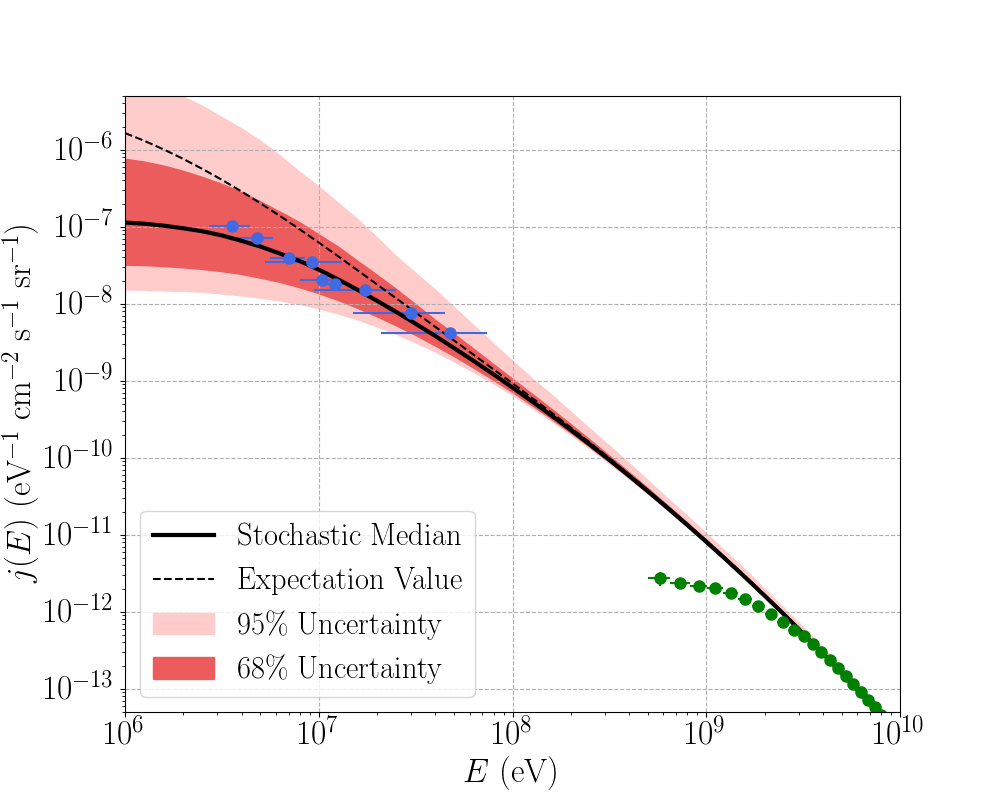}}
\caption{Stochastic fluctuations of GCR protons (left panel) and electrons (right panel) in comparison with data from Voyager~1~\cite{cummings2016} (blue) and AMS-02~\cite{AMS2014,AMS2015} (green) for the case of rigidity dependent diffusion coefficient. The dotted and solid black curves are respectively the expectation values and the median of the intensities. The shaded regions are the 95\% and 68\% uncertainty ranges.}
\label{fg:Drig}
\end{figure}

We note that the parameters in Tab. \ref{tab:parameters} have been obtained for the diffusion coefficient of the form as presented in the main text $D(E)\sim\beta \gamma^{\delta}$ which is expected when the magneto-static approximation is relaxed meaning the Alfv\'en speed is no longer negligible in comparison to the particle's speed in the resonance condition of wave-particle interaction (see e.g. \cite{schlickeiser1999,schlickeiser2010} for more technical details). In a broader sense, it is probably fair to admit that there remain significant uncertainties since there is currently no direct observations of the mean-free path in the interstellar medium. In order to bracket this uncertainty, we have also repeated our computation with a diffusion coefficient that has a power law dependence on rigidity, $D(E)\sim \beta R^\delta$ where $R$ is the particle's rigidity. We have found that this would not qualitatively change our results since the break in $D(E)$ below roughly 1 GeV does not significantly affect the spectrum of CR protons at low energies as the transport in this regime is dominated by energy loss. For CR electrons, the rigidity or Lorentz factor dependent diffusion coefficients are roughly the same down to 1 MeV. We present also in Fig.~\ref{fg:Drig} the fits for the case of a rigidity-dependent diffusion coefficient with slightly different values for the advection speed $u_0$ and the number density of hydrogen atoms $n_\mathrm{H}$ (see Tab.~\ref{tab:parameters2} for the complete list of parameter values in this case).

\begin{table}[h!]
\centering
\caption{Externally constrained and fitted parameters for the stochastic model for both CR protons and electrons  in the case for the diffusion coefficient scaling with rigidity.}

	\label{tab:parameters2}
	\begin{tabular}{|c|c|r|} 
		\hline
		\hline
		\multirow{2}{*}{\shortstack{Fitted parameters\\ for low-energy CRs}} & $n_{\mathrm{H}}$ & 0.7 cm$^{-3}$\\
		\cline{2-3}
		& $u_0$ & 18 km/s\\
		\hline
		\multirow{10}{*}{\shortstack{Constrained parameters\\ from high-energy CRs}} & $R_d$ & $\qquad$ 15 kpc \\
		\cline{2-3}
		& $H$ & 4 kpc\\
		\cline{2-3}
		& $2h$ & 300 pc\\
		\cline{2-3}
		& $2h_s$ & 80 pc\\
		\cline{2-3}
		& $D(E=10\textrm{ GeV})$ & $5\times 10^{28}$ cm$^2$/s\\
		\cline{2-3}
		& $\delta$ & 0.63\\
		\cline{2-3}
		& $\mathcal{R}_s$ & 0.03 yr$^{-1}$\\
		\cline{2-3}
		& $\xi_{CR}^{(p)}$ & 8.7\%\\
		\cline{2-3}
		& $\xi_{CR}^{(e)}$ & 0.55\%\\
		\cline{2-3}
		& $\alpha$ & 4.23\\   
		\hline
		\hline
	\end{tabular}
\end{table}

\bibliographystyle{apsrev4-2}
\bibliography{mybib}


\end{document}